# Astronomy from Coast to Coast to Coast

Eric Steinbring, *Eric.Steinbring@nrc-cnrc.gc.ca*
National Research Council Canada, Herzberg Astronomy and Astrophysics
5071 West Saanich Road, Victoria, B.C., Canada V9E 2E7

*This article is based on the Ruth Northcott Memorial Public Lecture given 1 July 2017, on the occasion of Canada Day 150 during the RASC GA in Ottawa.*

Canada is a triangle-shaped country, roughly speaking.  We all know that the Atlantic Ocean is at its eastern corner in the Maritimes, and off the west coast of British Columbia is the Pacific Ocean.  The Arctic Ocean, however, makes up the bulk of Canada's coastline, along its pointy "top."  That peaks closest to the North Pole in Nunavut, on the shores of Ellesmere Island.  Running down this island, as on our western flank in the Rockies, is a range of permanently snowcapped mountains, with one topping 2600 m.  It is a propitious geography, which along with that of northern Greenland, provides plenty of ice-locked, and windward, elevated coastal terrain; potentially perfect for astronomy. Actually, these mountains are about the same distance from either Halifax or Victoria (or Halifax to Victoria) as Victoria is from the amazing 4200-m summit of Maunakea, on the Big Island of Hawai'i – although they are not as easily reached.

Looking for the best location can make even the most arduous travels worthwhile though. Over the last decade, site testing beyond 80 deg North latitude on Ellesmere Island (near a place appropriately named "Eureka") has shown that Arctic coastal mountain sites can provide excellent astronomical sky conditions.[1]  It had already been well established over several decades that Antarctica provided great locations for astronomy, at latitudes similarly close to the South Pole, atop its roughly 3000-m high interior ice plateau.  A challenge for both places has been to overcome the difficulties of access.  In the middle of the icecap of central Antarctica, such as Dome C at 75 deg South latitude, scientists and engineers visit an incredibly harsh, effectively sterile place for a few months at a time, surrounded on all sides by hundreds of kilometers of glacier.

Canada's High Arctic is not easy to cross either, but it is alive.  There are flowers and birds, muskox and wolves, fish and seals.  And, of course, there are polar bears.  Although "arctic" may come from the Greek word arktos for bear, as in the celestial Ursa Major and Ursa Minor, it is the Inuit who have made this land home for many centuries, navigated with a separate tradition of watching and understanding the polar sky.  Nunavut today has hamlets this far north; communities with families, not just research stations.  Succeeding in the Far North still depends on finding your way, learning what tools have worked before, and adapting. This is true for precision astronomical instruments too, and I will present some examples that are illustrative.

## GOING TO EXTREMES

Despite being the most hostile place to live, space is ostensibly the ideal environment for astronomy. It is very cold, there is no atmosphere to obscure the view and the seeing is always perfect. There are some major caveats to that, of course. Beyond the cost to develop and launch a satellite, once deployed it is still potentially subject to a night/day or target-field duty cycle depending on orbit and spacecraft orientation, and it will have an inherently restricted operational lifetime. Balloon-borne missions might take advantage of the space-like conditions afforded by the stratosphere, but these too will have samples limited to something between a few hours up to weeks while that fragile balloon is aloft, and depending on both flight hardware and the vagaries of upper-atmospheric winds. From the ground, an alternative could be a large network of mid-latitude sites, although those will have another disadvantage: requiring their data to be laboriously stitched together under varying sky conditions and from different telescopes and instruments. For polar sites, extreme latitude instead ensures months-long windows of darkness, and combined with sufficient altitude this can offer a stable platform above much of the atmospheric water vapour, turbulence and cloud. A single location can give the same view, almost uninterrupted, for months. From the poles, targets have uniform airmass, each simply circling overhead and maintaining elevation.

An observer must get up fairly high to get above all the cloud though. In winter the High Arctic is usually blanketed in a strong atmospheric thermal inversion. Weirdly, it can be much colder at the surface than higher up. It might be -40 C at sea level, but "only" -25 C at 1000 metres above. But that helpfully keeps the clouds - mostly suspended ice crystals – below. Overhead is an isolated mass of cold air. Mountains on Ellesmere reach into this pristine region, but can be accessed by helicopter only in summer. Sadly, a helicopter cannot fly there in darkness. No matter how beautiful the view will be once the Sun goes down, a person cannot safely remain. The trick is to install a robot that can.

## LEARNING FROM THE EXPERTS

There is plenty of existing know-how to make this work. Military infrastructure for Alert, the northernmost outpost on Ellesmere Island, is part of that legacy. A series of microwave-repeater stations stretches down to Eureka, which is the furthest north from which one can still see a geosynchronous communications satellite. This is the wireless "backbone" of Ellesmere Island, and it is powered by batteries and serviced by helicopter. Many other similar, but smaller, examples are a host of autonomous weather stations at various locations in the Arctic, and elsewhere. These too are visited, typically, just once a year to replenish batteries, download data, and fix broken parts.  Many of these geophysical stations are operated by government and university-led groups, and their fieldwork is supported - complete with helicopters, bushplanes, tents and cookstoves - by the Polar Continental Shelf Program facility from Resolute Bay, at 75 deg North.

That model was followed for the deployment of three stations on the northernmost isolated Ellesmere Island mountains, between 1100 m and 1900 m  elevation, closest to the coast, and that were not in the restricted region of Quittinirpaaq National Park.  These were nicknamed "Inuksuit" like the stone waymarkers of the North.[2]  Unlike the iconic inuksuk, however, each was equipped with an autonomous weather station, a camera, a satellite antenna and a wind turbine.  A later addition was "Ukpik", which in addition to a fuel cell for generating electricity, had a 3-cm aperture telescope for continuously monitoring Polaris.[3]  An all-sky-viewing camera made the analogy to its Inuktitut namesame of "snowy owl" even closer, as its smooth, white cowling looked a lot like one hunkered down in the snow.  And better yet, when confronted with strong winds and blowing snow - or daylight - a protective cover would slide over both of those.  Ukpik was asleep.  These systems completed their work after 5 years, and were removed. Weather and sky clarity measurements nicely matched predictions, with some evidence that seeing is good.  The necessary logistics were demonstrated.  But the complexity of those logistics is not trivial, nor is the cost.

Otherwise, a compromise is closer to Eureka, a facility called the Polar Environment Atmospheric Research Laboratory (PEARL).  Again, that is an apt name, as this unique shiny "pearl" is rather precious, sitting all by itself on a 600 m-high ridge.  It can be accessed by large 4-by-4 diesel pickups along a winding 15 km-long gravel road from the sea-level base at Eureka.  That is operated by the federal weather service of Environment and Climate Change Canada, with a year-round rotating crew of eight.  Eureka is a homey research station with real beds, proper meals, a jet-capable airstrip, and a yearly resupply ship.  PEARL is a practical observatory, designed for continuous atmospheric studies with optical instruments, with a large flat observing platform on its roof.  This has supported various astronomical observations, with much thanks to a university-led consortium known as the Canadian Network for Detection of Atmospheric Change, which maintains many other instruments there, and has lots of experience doing so.

From those atmospheric-science instruments we know some of the good basic qualities of PEARL.  All-sky-monitoring camera images taken over many winters showed half were obtained under perfectly photometric conditions (no cloud at all),  about two-thirds were clear (no clouds visible to the unaided eye) for periods of up to 100 hours at a time, with just over 15% "unusable", so more than 2 mag of extinction in V band.[4]  That is about the same as Maunakea, among the best optical/infrared observatory sites.  And the sky is similarly dark in the visible: down below 20 mag per square-arcsec.  But here is where the -30 C and colder conditions really start to show their advantage, which can be teased out from years of archival spectrometer data.  In the thermal infrared, past 2.4 micron wavelength, the sky emits something like only 18 mag per square-arcsec, similar to South Pole.  But this is 10 to 20 times darker than what is typical for Maunakea, which usually dips just below freezing at night, and will be even warmer during the day. [5]

The PEARL facility has much easier access than the high Ellesmere Island mountains.  You can arrive in Eureka via a large aircraft chartered from Yellowknife, in the Northwest Territories, during dead of winter.  Best of all, observations can be in campaign-mode, meaning that a person will be on site to operate the instrument.  When something breaks, or some snow accumulates on a lens, they can go and fix the problem.  Maybe that just means going out on the roof with a broom.  But the value of that should not be underestimated.  Although there is not much precipitation here year-round, and precious little snow in winter, winds can make troublesome snowdrifts.  This becomes particularly evident in the humbling situation when the observer gets their big 4-by-4 truck stuck in a mere two-foot-high snowdrift on the way down from the summit.  Snowpack here is bone dry and very hard.  It is much more styrofoam-like than the fluffy snowball-making snow we southern Canadians are more familiar with.  I was assured that even the most seasoned drivers get stuck once in a while, but still embarrassed the one time a heavy-equipment operator had to come up with a front-end loader to help me.

### SEEING A WAY FORWARD

An important advantage of PEARL is that standard site-testing instruments employed elsewhere in the world, often using commercial Cassegrain telescopes of about 30-cm aperture, can be used.   Granted, these are not quite off-the-shelf solutions.  Mounts and electronics must be modified to operate under stricter conditions.  Also, the usual method of continuously heating a Schmidt plate to keep it clear of dew or frost is best not done if you want to look at Polaris to measure seeing.  To do so you always point straight overhead, essentially.  That was the job of "Ukaliq", as in the arctic hare. [6]  Just like the well-adapted hare, which sometimes gets up on its hind legs to get a better view, Ukaliq tips up only when observing.  Of course, under calm observing conditions ice crystals precipitate nearly straight down and stick to the telescope corrector plate.  Worse, accumulated melted crystals will pool.  It can be better to turn the telescope down-facing at intervals and gently blow on it with a fan to sublimate frost, which is also easily automated.  Or simply wipe it off with some alcohol.  The glass is a lot easier to reach this way than when it is well above head height and pointed straight up.

Deployment of this and other instruments have shown how astronomy from Canada's northern tip can fit into the worldwide picture.  We know the seeing at PEARL is very good, in part due to a clever device called the Arctic Turbulence Profiler (ATP), a lunar scintillometer.[7]   The full Moon will be up in winter, and then conveniently stay at the same low elevation the whole time, which is perfect if you want to watch it to monitor scintillation.   Data from the ATP, and from measurements of binary stars with another specialized instrument called a Slope Detection and Ranging (SloDAR) device have confirmed that most deterioration takes place close to the ground.[8]  The part of seeing associated with just the upper atmosphere is typically 0.23 arcsec full-width at half-maximum.  Half the time, total seeing measured from 8-m elevation is 0.76 arcseconds or better.[9]  Again, these numbers are comparable to the best mid-latitude sites.  Autonomous cameras have also shown that high-precision, millimag-level photometric accuracy is possible from PEARL, just as has been done already from Dome A (80 deg South latitude) by the Chinese Small Telescope ARray (CSTAR), for example.  Like those, the Arctic Wide-field Cameras (AWCams) were meant for variable-star and exoplanet transit detections with small, roughly 10-cm apertures.[10]  That is limited primarily by scintillation, and takes particular advantage of wide, uninterrupted clear swaths of sky.  Cold-hardening larger, pointable telescopes is a bigger challenge that can be overcome, and justified by strong scientific cases that require the special conditions only the Poles provide.[11]  The history here has been one of choosing the right path and tools by sharing knowledge, learning from experience, and adjusting course when needed.

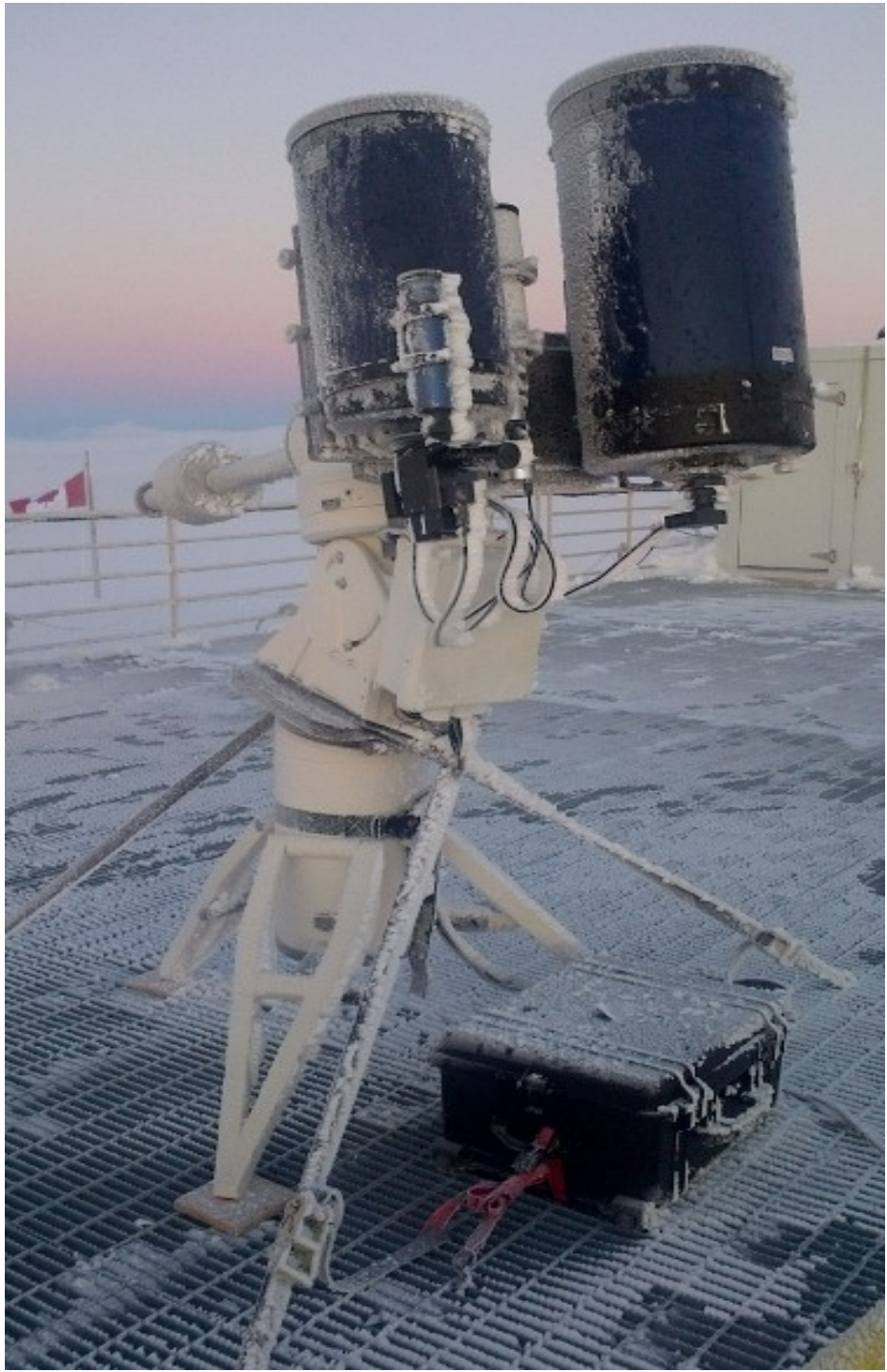

The Ukaliq autonomous site-testing telescope setup on the PEARL rooftop (left) seen dusted with frost, just at sunrise after one long "night" of observing Polaris.